\documentclass[5p]{elsarticle}
\usepackage[utf8]{inputenc}
\usepackage{graphicx} % Required for inserting images
\usepackage{siunitx}
\usepackage[colorlinks = true,linkcolor = blue,urlcolor  = blue, citecolor = blue]{hyperref}
\usepackage{breakurl}
\usepackage{amsmath}
\usepackage{lipsum}
\usepackage{subcaption}
\usepackage[version=4]{mhchem}
\usepackage{overpic}
\usepackage[capitalise]{cleveref}
\usepackage{glossaries} 
\usepackage{supertabular,booktabs}

\makeglossaries

\glsaddkey{unit}{\glsentrytext{\glslabel}}{\glsentryunit}{\GLsentryunit}{\glsunit}{\Glsunit}{\GLSunit}

\newglossarystyle{symbunitlong}{%
\setglossarystyle{long3col}% base this style on the list style
  {\end{supertabular}}%

}

\newacronym{fpp}{FPP}{Fusion Power Plant}
\newacronym{dt}{D-T}{Deuterium-Tritium}
\newacronym{tbr}{TBR}{Tritium Breeding Ratio}
\newacronym{lib}{LIB}{Liquid Immersion Blanket}
\newacronym{libra}{LIBRA}{Liquid Immersion Blanket: Robust Accountancy}
\newacronym{baby}{BABY}{Build A Better Yield blanket}
\newacronym{lsc}{LSC}{Liquid Scintillation Counting}

\journal{Nuclear Fusion}

\begin{document}
\begin{frontmatter}
\title{Advancing Tritium Self-Sufficiency in Fusion Power Plants: Insights from the BABY Experiment}
\cortext[mycorrespondingauthor]{Corresponding author}
\ead{remidm@mit.edu}

\author[MIT]{Rémi Delaporte-Mathurin\texorpdfstring{\corref{mycorrespondingauthor}}}
\author[MIT]{Nikola Goles}
\author[MIT]{John Ball}
\author[MIT]{Collin Dunn}
\author[MIT]{Emily Edwards}
\author[MIT]{Sara Ferry}
\author[EHS]{Edward Lamere}
\author[MIT]{Andrew Lanzrath}
\author[MIT]{Rick Leccacorvi}
\author[MIT]{Samuele Meschini}
\author[MIT]{Ethan Peterson}
\author[MIT]{Stefano Segantin}
\author[MIT]{Rui Vieira}
\author[MIT]{Dennis Whyte}
\author[MIT]{Weiyue Zhou}
\author[MIT]{Kevin Woller}

\address[MIT]{Plasma Science and Fusion Center, Massachusetts Institute of Technology, Cambridge, MA 02139, USA}
\address[EHS]{Environment, Health \& Safety Office, MIT}

\begin{abstract}
In the pursuit of fusion power, achieving tritium self-sufficiency stands as a pivotal challenge.
Tritium breeding within molten salts is a critical aspect of next-generation fusion reactors, yet experimental measurements of \gls{tbr} have remained elusive.
Here we present the results of the \gls{baby} experiment, which represents a pioneering effort in tritium research by utilizing high-energy (\SI{14}{\mega\electronvolt}) neutron irradiation of molten salts, a departure from conventional low-energy neutron approaches. 
Using a small-scale (\SI{100}{\milli\litre}) molten salt tritium breeding setup, we not only simulated, but also directly measured a \gls{tbr}. 
This innovative approach provides crucial experimental validation, offering insights unattainable through simulation alone.
Moreover, our findings reveal a surprising outcome: tritium was predominantly collected as HT, contrary to the expected TF. 
This underscores the complexity of tritium behavior in molten salts, highlighting the need for further investigation.
This work lays the foundation for a more sophisticated experimental setup, including increasing the volume of the breeder, enhancing neutron detection, and refining tritium collection systems.
Such improvements are crucial for advancing our understanding of fusion reactor feasibility and paving the way for future experiments.

\end{abstract}

\end{frontmatter}

\printglossary[type=\acronymtype]
\glsresetall

\section{Introduction}

Most proposed \glspl{fpp} are fueled by  \gls{dt} fusion reactions \cite{fia2023}, but tritium is scarce \cite{kovari_tritium_2017, pearson_tritium_2018}.
It is essential to develop tritium breeding blanket and fuel cycle technologies that can generate and maintain a self-sustaining supply of tritium within the \gls{fpp}.
Tritium self-sufficiency refers to the ability of the \gls{fpp} to breed and process enough tritium to run the plant without relying on an external source \cite{coleman_demo_2019, chen_tritium_2016, meschini_modeling_2023, abdou_physics_2021}.
Achievement of tritium self-sufficiency is intrinsically tied to the design and performance of the tritium breeding blanket.

However, tritium self-sufficiency on a large scale has never been demonstrated.
The \gls{libra} project was proposed at the MIT Plasma Science and Fusion Center (PSFC) to address critical research gaps in understanding tritium breeding and tritium chemistry in molten FLiBe exposed to a fusion neutron environment \cite{ferry_libra_2022}.
FLiBe has been proposed as a breeder material for the \gls{lib} of ARC-class \glspl{fpp} \cite{sorbom_arc_2015}.
In an ARC-class \gls{fpp}, the first wall and vacuum vessel structures are coupled and located inside a tank of molten FLiBe salt.
The \gls{lib} is a self-cooled blanket concept \cite{kondo_606_2020}, and the FLiBe serves as breeder, neutron shield, and coolant for the vacuum vessel \cite{kuang_conceptual_2018}.
The beryllium in FLiBe acts as a neutron multiplier, which significantly enhances \gls{tbr}.
Although there are many challenges associated with working with beryllium because of its toxicity, FLiBe remains the most promising breeder choice for the ARC-class \gls{fpp} due to its cooling capabilities \cite{ferrero2023exploration}, and its ability to achieve high \gls{tbr} \cite{segantin_optimization_2020}.
Recent models have shown that the FLiBe \gls{lib} should enable the ARC-class \gls{fpp} to achieve tritium self-sufficiency \cite{meschini_modeling_2023}.

The primary parameter used to describe the efficiency of a breeding blanket is the \gls{tbr}.
The \gls{tbr} is defined as the ratio of tritium produced in the blanket to tritium consumed in the plasma.
To achieve self-sufficiency, the \gls{tbr} of a commercial fusion reactor must be greater than unity, meaning that more tritium is generated than consumed during the fusion process.
The \gls{tbr} required by a given \gls{fpp} design depends on the plant's performance requirements, the efficiency of its fuel cycle components, expected tritium losses, and reserve inventory needs.
Designing a breeding blanket with a \gls{tbr} high enough is a critical engineering challenge on the path to commercial fusion power.

The long-term goal of \gls{libra} is to demonstrate a ${\mathrm{TBR} \geq 1}$ in a large volume (\SI{1000}{kg} $\sim$ \SI{500}{L}) of FLiBe molten salt using \gls{dt} neutron generators.
Note that a full-scale \gls{lib} in an ARC-class \gls{fpp} will require $\sim$ \SI{250,000}{L} of FLiBe, hence the importance of understanding tritium behavior in large salt volumes.
\gls{libra} presents major safety and infrastructure challenges related to beryllium, high-energy neutrons, tritium, high-temperature fluids ($>$ \SI{750}{K}) and hydrofluoric acid formation.
Therefore, the PSFC is undertaking a scaled approach to develop the knowledge, team, and techniques necessary to achieve the goals of \gls{libra}, beginning with the \gls{baby} campaign, which is the focus of this paper.

The ITER and DEMO research programs have adopted a similar scaled approach through the test blanket module program \cite{giancarli2006breeding}, which evaluates fabrication, cooling, and tritium production capabilities using blanket mock-ups for the Helium Cooled Lithium Lead (HCLL), Helium Cooled Pebble Bed (HCPB), Helium Cooled Ceramic Breeder (HCCB), and Water Cooled Ceramic Breeder (WCCB) concepts \cite{forest2020test, wang2019current, hirose2024functional}.
Neutronics experiments were conducted on the mock-ups with \SI{14}{MeV} neutrons at low neutron fluxes ($<$\SI{5e8}{n.cm^{-2}.s^{-1}} ), aimed at validating numerical tools, developing neutron detection techniques, and quantifying tritium production rates \cite{batistoni2012neutronics, ochiai2014dt}.
Similar neutronics experiments were carried out for the WCCB blanket of the China fusion engineering test reactor (CFETR) \cite{zhu2021experimental}.

%Geometry constraints and performance requirements on the breeding blanket are specific to the \gls{fpp} design.

This paper presents the first results of the \gls{baby} experiment.
\gls{baby} is a small-scale (\SI{100}{\milli\liter}) molten FLiBe salt tritium breeding experiment that uses \SI{14}{MeV} \gls{dt} neutrons and is the initial experimental step towards the construction and operation of \gls{libra}.
Smaller-scale experiments enable the \gls{libra} team to gain experience in neutron generation, neutron detection, and tritium accountancy and refine the experimental plans for \gls{libra} while minimizing safety hazards.
Furthermore, it enables the near-term generation of useful new data on tritium breeding in molten salts.
For the initial test results presented here, the chemical hazard associated with FLiBe was alleviated by using a surrogate salt: ClLiF, a mixture of lithium fluoride (LiF) and lithium chloride (LiCl).
By avoiding the presence of beryllium for the initial tests, the team was able to build and troubleshoot the \gls{baby} experimental setup before introducing the additional safety protocols and equipment required for beryllium work.
    
\section{Principle}\label{principle}

\begin{figure*}[h]
    \centering
    \begin{subfigure}[b]{0.7\textwidth}
        \centering
        \includegraphics[width=\linewidth]{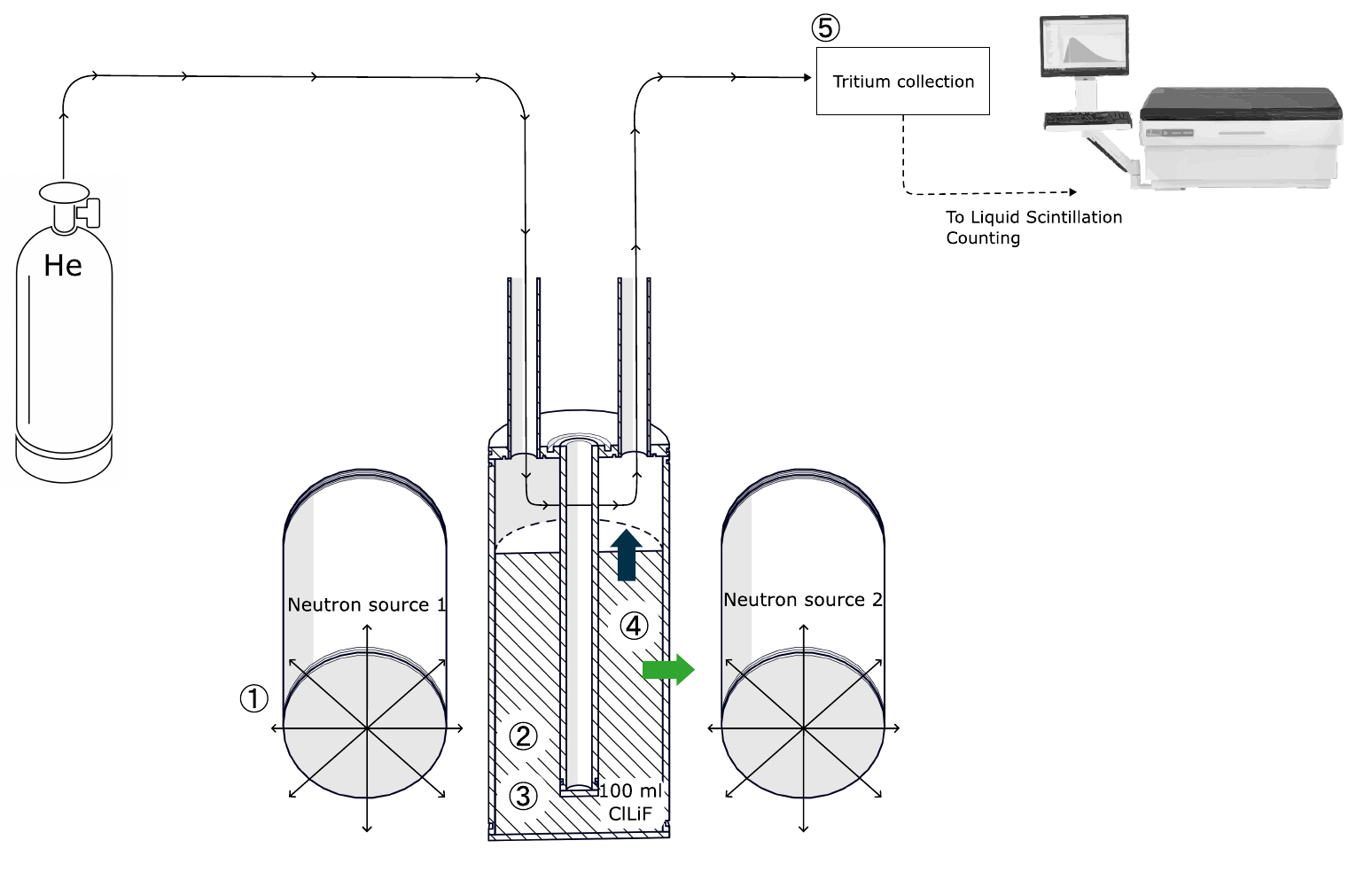}
        \caption{Experimental setup. The thermal insulation around the crucible is not shown here.}
        \label{fig:experiment principle}
    \end{subfigure}
    \par\bigskip
    \begin{subfigure}[b]{0.5\textwidth}
        \includegraphics[width=1\linewidth]{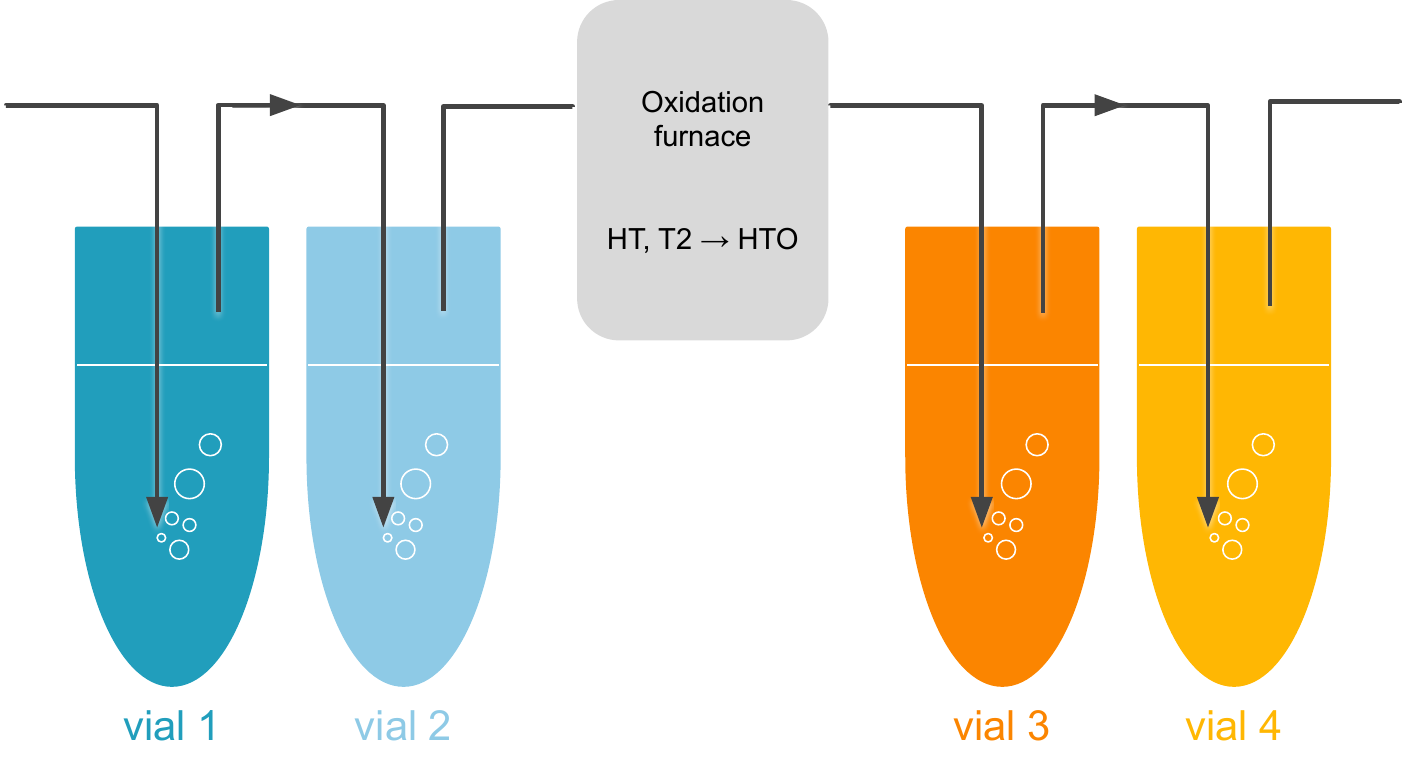}
        \caption{Tritium detection principle. The first two vials (in blue) collect tritium in soluble form (HTO, TCl and TF). Insoluble tritium (HT and T$_2$) is then passed through an oxidation furnace, where it is converted into HTO and collected in the second pair of vials (in orange).}
        \label{fig:tritium detection principle}
    \end{subfigure}
    \caption{\gls{baby} experimental setup.}
\end{figure*}

This section describes the general principle of the \gls{baby} experiment.
Higher level methodology details are available in \cref{methods}.

The \gls{baby} experiment is composed of a crucible containing \SI{100}{\milli\liter} of molten salt, two sources of \SI{14}{MeV} neutrons, a gas system, a neutron detection system and a tritium collection system (see \cref{fig:experiment principle}):

\begin{enumerate}
    \setcounter{enumi}{-1}
    \item The system is conditioned with \SI{3.5}{\percent} H$_2$ in He for \SI{4.5}{h} at \SI{30}{ccm} followed by at least \SI{1}{h} of pure He at \SI{20}{ccm}. The salt is then brought to \SI{700}{\celsius}. It takes a few hours for the salt temperature to equilibrate.
    \item Once the system is at equilibrium, \SI{14}{MeV} neutrons are generated.
    \item Neutrons interact with lithium in the salt to produce tritium according to the following reactions: \begin{align}
            \ce{^6Li + n &-> He + T + \SI{4.8}{MeV}}\\
            \ce{^7Li + n &-> He + T + n$'$ - \SI{2.5}{MeV}}
        \end{align}
    \item Tritium is transported in the salt through diffusion and / or advection due to natural convection currents in the molten salt \cite{humrickhouse_tritium_2020}.
    \item Tritium eventually reaches the top surface of the molten salt, which is in contact with the sweep gas, and the tritium is released into the gas stream, or reaches the salt-metal interface at the walls of the crucible, permeates the metal wall, and is released into the room. 
    \item Tritium in the gas stream is collected and counted.
\end{enumerate}

The ratio of the convective mass transfer to the diffusive mass transfer is expressed by the Sherwood number:
\begin{equation}
    \mathrm{Sh} = \frac{k}{D/L}
    \label{eq:sherwood}
\end{equation}
where $k$ is the mass transport coefficient (\si{\metre\per\second}), $D$ is the diffusion coefficient (\si{\metre\squared\per\second}), and $L$ is the characteristic length of the system (\si{m}).
When $\mathrm{Sh} \gg 1$ the dominant mode of mass transfer is advection.
When $\mathrm{Sh} \ll 1$, the dominant mode is diffusion.
In the \gls{baby} experiment, it is expected that $\mathrm{Sh} \gg 1$.

\subsection{Molten salt crucible}

\SI{190}{\gram} of molten ClLiF salt, with stochiometric composition 30.5\% LiF - 69.5\% LiCl (eutectic), were placed in a \SI{1.6}{mm} thick Inconel-625 crucible (see \cref{fig:experiment principle}).
The internal height of the crucible is \SI{101}{mm} and its internal diameter is \SI{42}{mm}.
Detailed drawings and CAD models are available as supplementary data.
Helium was swept across the top surface of the ClLiF inside the crucible to carry away the released tritium.
The molten salt was kept at \SI{700}{\celsius} using a re-entrant heater in the center of the crucible.
The crucible was wrapped in alumina wool for thermal insulation.
The heater was connected by a temperature controller and a sheathed thermocouple immersed in the salt.

\subsection{Neutron source} \label{neutronsource}
 A Thermo-Fisher\textsuperscript{TM} A-325 and a Thermo-Fischer\textsuperscript{TM} P-383, both sealed-tube neutron generators, were used for the \gls{baby} experiment.
%A sealed tube neutron generator generates \SI{14}{MeV} neutrons by D-T fusion reactions \cite{ludewigt_high-yield_2007, verbeke_development_2000}.
Sealed-tube neutron generators house a deuterated and tritiated metal target that is surrounded by a high voltage electrode.
High voltage is applied across the electrode, creating an electric field that accelerates deuterium ions into the target, where they slow down due to Coulomb collisions.
A fraction of deuterium ions undergo \gls{dt} nuclear fusion reactions, producing a He nucleus and a \SI{14}{MeV} neutron.
The sealed design ensures that the radioactive tritium fuel remains contained, and the generated neutrons can be employed for various applications.

\subsection{Neutron detection}

The neutron detection system on the \gls{baby} experiment is a combination of activation foils and diamond detectors.

% Activation foil analysis
Activation foil analysis is a neutron detection technique based on the principle of inducing nuclear reactions in target materials known as activation foils \cite{greenberg_neutron_2011}.
The populations of radioactive isotopes created in the foil are characterized by using a suitable detection system (e.g. gamma ray spectroscopy), and these can be directly correlated to the energy spectrum and fluence of the incident neutrons.
Foils can be strategically placed to characterize the neutron flux at different locations of interest.

% Diamond detectors
Diamond detectors were used to measure the neutron flux near the crucible.
Diamond detectors operate on the basis of the neutron-induced displacement damage process in the crystalline lattice of a diamond \cite{angelone_properties_2021}.
The resulting vacancy and interstitial defects act as electrically active centers within the diamond structure and create a measurable electrical signal that can be correlated to the characteristics of the impinging neutron flux.
The most common type of diamond detector for neutron detection utilizes synthetic diamond doped with isotopic impurities such as nitrogen-vacancy (NV) centers.
The NV centers serve as sensitive charge carriers and enhance the detection efficiency.
Diamond detectors exhibit high sensitivity to both thermal and fast neutrons, making them versatile tools for neutron spectroscopy.

\subsection{Tritium detection}

The tritium detection method is \gls{lsc} \cite{parker_radiometric_2023}.
The gas containing traces of tritium first passes through a volume of water, stripping out all soluble forms of tritium (HTO and TF) (see \cref{fig:tritium detection principle}).
It is then passed through an oxidizing furnace, which converts insoluble forms of tritium (HT, T$_2$) into HTO.
The gas is then passed through another volume of water to collect the remaining tritium.
Therefore, it is possible to discriminate between the soluble and insoluble forms of tritium.
The radioactivity of tritiated water is then measured by \gls{lsc}.
\section{Results}

All data, analysis and post-processing scripts are available at \url{https://github.com/LIBRA-project/BABY-paper} \cite{delaporte-mathurin_libra-projectinsights--baby-experiment-paper_2024}.

\subsection{\gls{tbr} measurement}\label{tbr_measurement}

\begin{figure}[h]
     \centering
     \begin{subfigure}[b]{\linewidth}
         \centering
         \includegraphics[width=\linewidth]{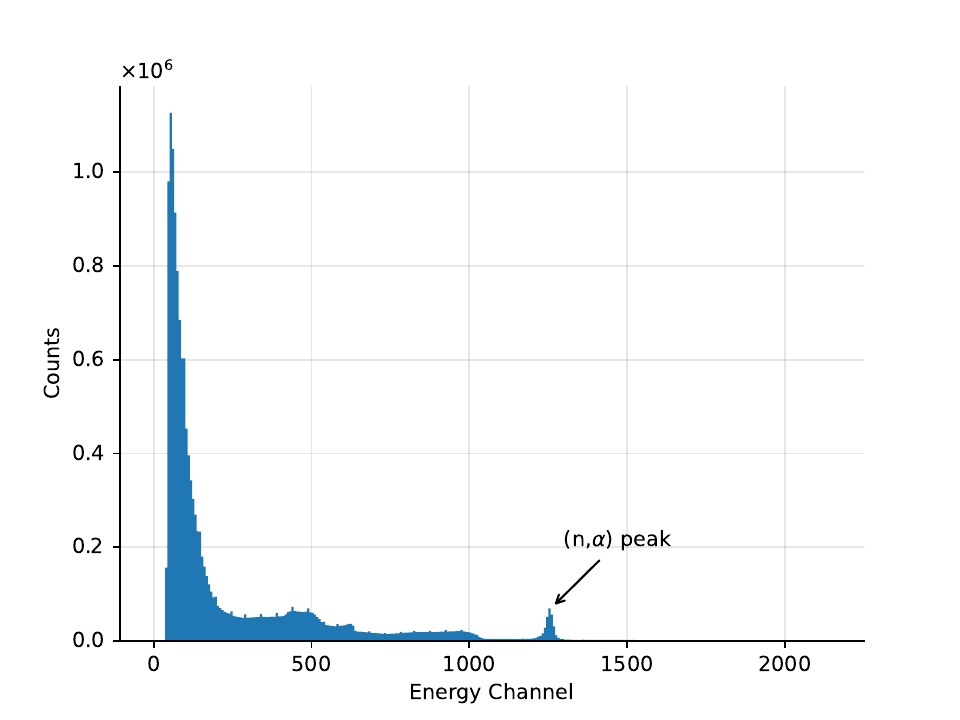}
         \caption{Combined energy spectrum}
         \label{fig:combined energy spectrum}
     \end{subfigure}
     \begin{subfigure}[b]{\linewidth}
         \centering
         \includegraphics[width=\linewidth]{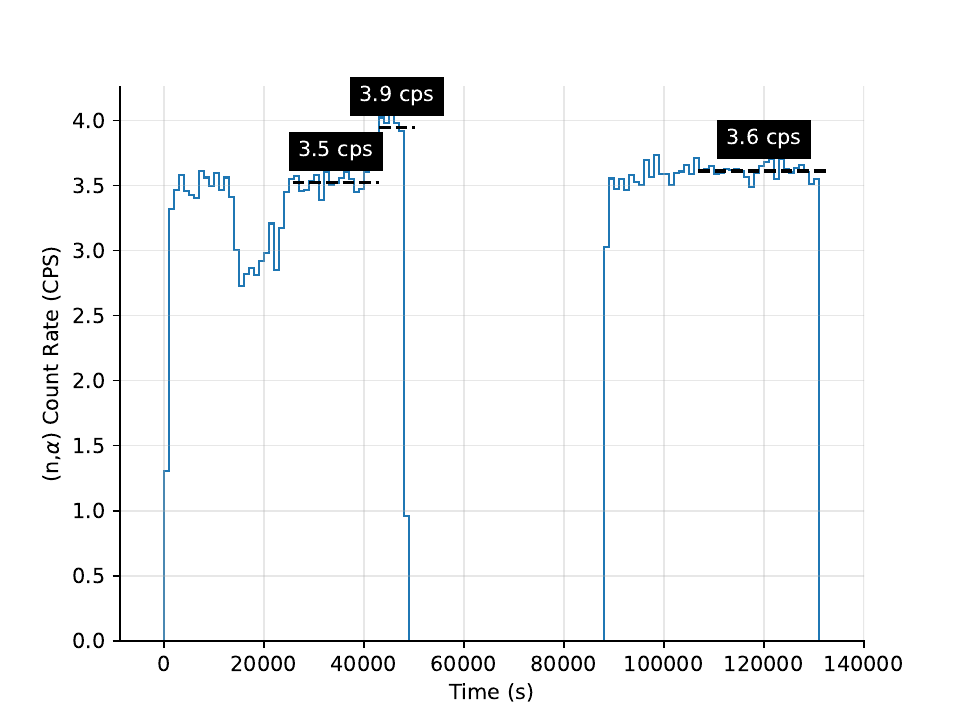}
         \caption{(n,$\alpha$) peak count rate}
         \label{fig:n alpha peak count rate}
     \end{subfigure}
    \caption{Diamond neutron detector results during BABY irradiation experiment.}
    \label{fig:neutron detection results}
\end{figure}

\paragraph{Neutron fluence measurement}
The diamond detector was used to obtain a real-time measurement of the neutron count rate. %; however, it was not employed for the full extent of the irradiation schedule until run \#5.
The $(\mathrm{n}, \alpha)$ peak can be seen between channel 1180 and 1300 of the uncalibrated neutron energy spectrum (see \cref{fig:combined energy spectrum}).
Counts within these energy channels corresponding to the $(\mathrm{n}, \alpha)$ peak were binned with a bin size of 1000 seconds (see \cref{fig:n alpha peak count rate}).
During most of the irradiation, this peak count rate remained steady at about 3.5 to 3.6 counts per second (CPS), but during the first irradiation period, a problem occurred with the P383 neutron generator, resulting in the generator being turned on and off and tuned repeatedly, explaining the dip seen between $t=\SI{16000}{s}$ and $t=\SI{25000}{s}$ in \cref{fig:n alpha peak count rate}.
Additionally, for an unknown reason, the count rate increased to an average rate of 3.9 CPS between $t=\SI{43000}{s}$ and $t=\SI{50000}{s}$.
During the second irradiation (\qtyrange{90000}{133000}{s}), the peak count rate remained constant at 3.6 CPS.

The results plotted in \cref{fig:neutron detection results} represent the count rate of the detector pulses from the diamond detector, but to convert these results into an actual neutron rate requires a measurement of the intrinsic efficiency of the diamond detector.
This measurement is planned for future experiments.

However, by also employing activation foil analysis, the total neutron fluence and therefore the average neutron rate were measurable.
Niobium activation foils were placed on top of each generator and aligned with the target plane.
The neutron rates of the P383 and A325 generators were calculated to be \SI{2.71e8}{n\per \second{}} and \SI{1.16e8}{n\per \second{}}, respectively, with a relative error of \SI{8.9}{\percent}.
The total neutron fluence over \SI{24}{\hour} of irradiation is therefore \SI[separate-uncertainty = true]{3.35(0.30)e13}{n}.

\paragraph{Tritium production measurement}

\SI{21}{\becquerel} of tritium\footnote{For convenience, tritium quantities are given in \si{\becquerel} based on the tritium specific activity \SI{3.57e14}{\becquerel\per\gram}.} were collected in total (see \cref{fig:tritium results}).
The collection vials were changed every 12 hours before day 2 and then regularly until no more tritium was collected.
The tritium in the water was then counted with the \gls{lsc} .
Based on the collected amounts, the overall collection efficiency (with two vials) was calculated to be $>\SI{99.7}{\percent}$.

Because each neutron is the product of one triton being consumed in a \gls{dt} reaction in the generators, the \gls{tbr} of the experiment can be calculated from the ratio of total tritium production in \gls{baby} to measured neutron fluence:

\begin{equation}
    \mathrm{TBR} = \frac{\SI{1.17e10}{T}}{\SI{3.35e13}{n}} = \num[]{3.57e-04}
\end{equation}

% OpeMC calculation of TBR
\paragraph{Modeling TBR}

\begin{figure*}[h]
     \centering
     \begin{subfigure}[b]{0.7\textwidth}
         \centering
        \includegraphics[width=\linewidth]{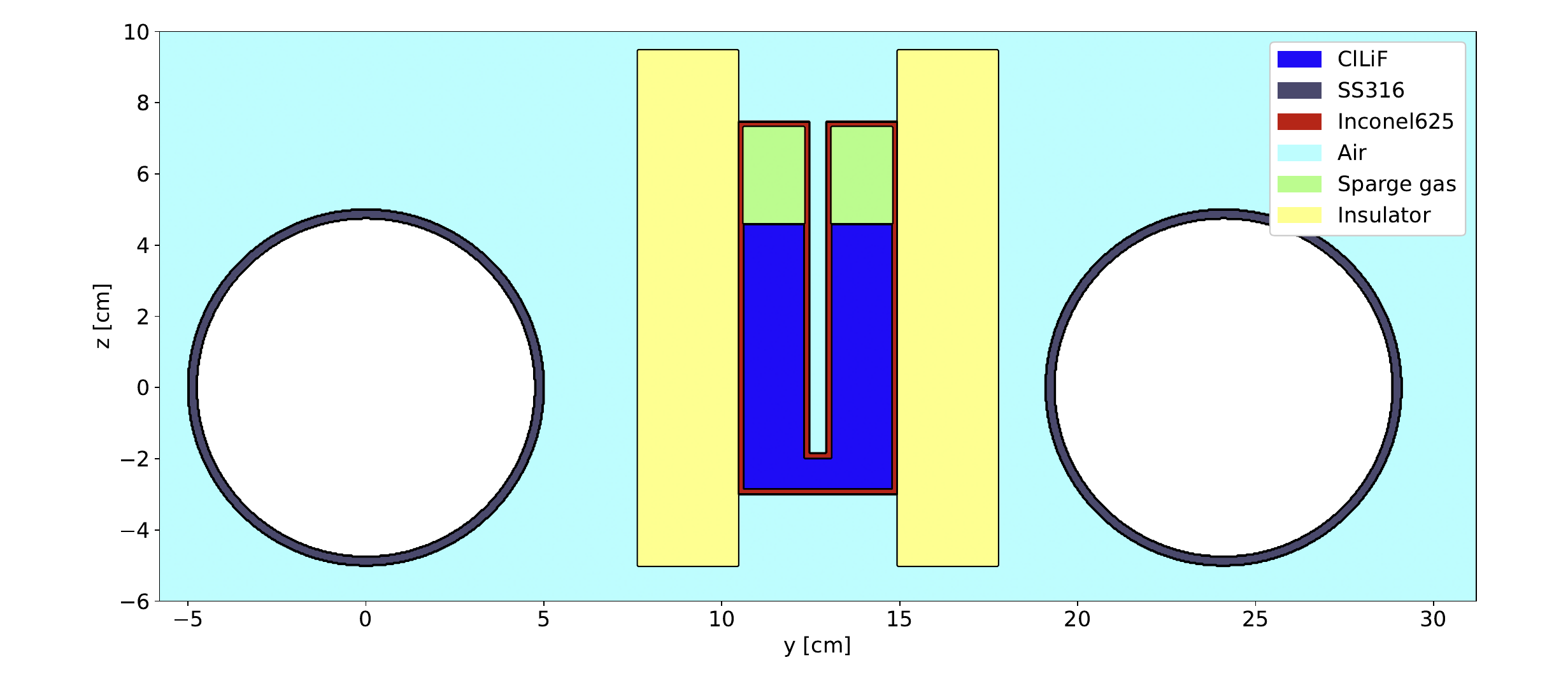}
        \caption{Geometry used to model the \gls{baby} experiment in OpenMC.}
         \label{fig:openmc-model}
     \end{subfigure}
     \begin{subfigure}[b]{0.49\textwidth}
         \centering
         \includegraphics[width=\linewidth]{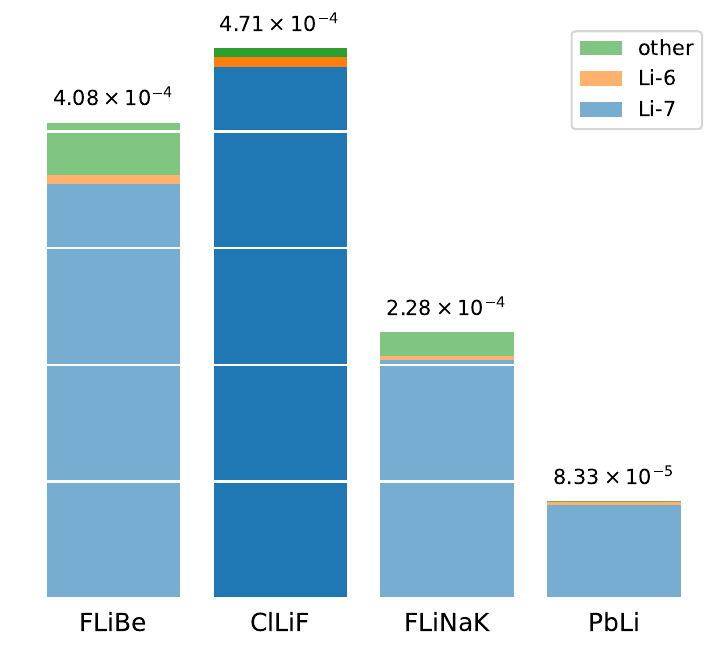}
        \caption{\gls{tbr} values for FLiBe, ClLiF, FLiNaK, and PbLi from OpenMC simulations of the \gls{baby} experiment. The contribution of lithium isotopes to tritium production is shown.}
        \label{fig:openmc-tbr}
     \end{subfigure}%
     \hfill
     \begin{subfigure}[b]{0.49\textwidth}
         \centering
            \includegraphics[width=\linewidth]{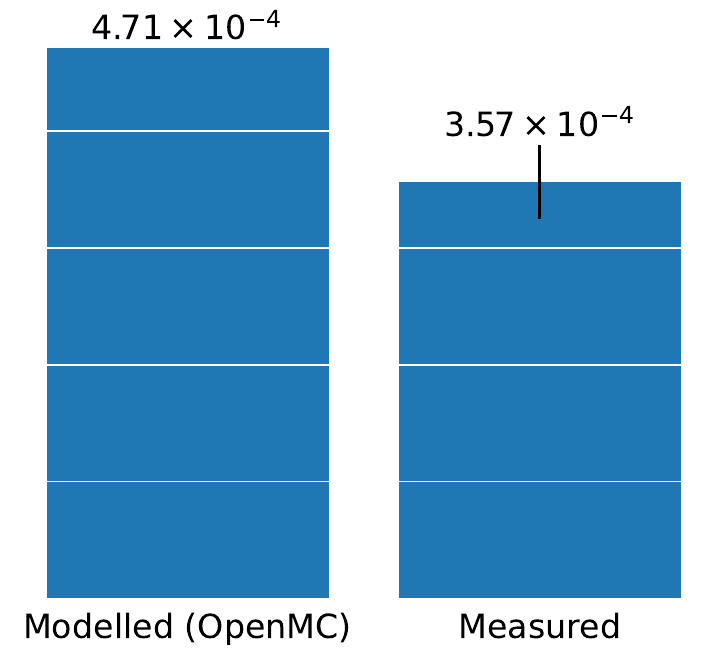}
            \caption{Comparison between the experimentally measured \gls{tbr} for \gls{baby} using ClLiF salt and the \gls{tbr} calculated in the corresponding OpenMC simulation.}
            \label{fig:comparison modelled measured tbr}
     \end{subfigure}
    \caption{Results of the OpenMC model. The reported values represent the statistical mean derived from Monte Carlo evaluations, with a relative standard deviation below 0.1\%.}
    \label{fig:three graphs}
\end{figure*}

% The experiment was formulated through iterative neutronics modeling using the OpenMC code \cite{romano_openmc_2015}.
% The primary objective was to optimize the experimental configuration to maximize the Tritium Breeding Ratio (TBR).
% Given the experiment's scale, there was a concern that the produced tritium might be too low to be detectable.
% The determined \gls{tbr} was adequate to establish acceptable irradiation schedules, ensuring a detectable amount of tritium.
% Furthermore, a validated \gls{tbr} value was crucial for the tritium release model, with neutron detection results used for model validation. The OpenMC model is illustrated in  \cref{fig:openmc-model}.

The experiment was modeled using OpenMC \cite{romano_openmc_2015}.
The model geometry included the neutron sources, crucible, thermal insulation, salt, and different gases (see \cref{fig:openmc-model}).
modeling the crucible and the two sources was deemed sufficient for a robust \gls{tbr} evaluation.
A parametric study was carried out to maximize the \gls{tbr} achievable in \gls{baby}.
This study considered parameters such as the position of the source relative to the crucible, as well as the implementation of moderators and reflectors.
The study showed that minimizing the source distance from the crucible was the most effective parameter for increasing the \gls{tbr}, as this increases the total fluence of neutrons in the salt.
The modeled \gls{tbr} was sufficiently high that the inclusion of reflectors and moderators were deemed unnecessary.
During the experiment, the two neutron sources were positioned as close to the crucible as was possible without exceeding their maximum operating temperatures.
The same model was applied to test and compare the \gls{tbr} of different breeders at a \SI{100}{\milli\liter} scale.
The OpenMC study included FLiBe, which is the tritium breeder of choice for the ARC reactor concept \cite{sorbom_arc_2015, kuang_conceptual_2018}, PbLi - which is considered for some DEMO reactor breeding concepts \cite{kondo_606_2020}, as well as FLiNaK and ClLiF, which are lithium-based molten salts that do not contain beryllium.
They are easier to handle than FLiBe and can breed tritium, although their breeding potential is lower than FLiBe, since beryllium acts as a neutron multiplier.

At this scale, \gls{tbr} is about four orders of magnitude lower than the value required by commercial fusion reactors (see \cref{fig:openmc-tbr}).
However, with the irradiation schedule and tritium detection systems foreseen for this experiment, a \gls{tbr} on the order of \num{e-4} was considered just sufficient to detect tritium.
The TBR is low because only 
 \SI{2}{\percent} of the neutron generator flux is incident on the salt (corresponding to a solid angle $\approx\SI{0.3}{\steradian}$).
 Future experiments in the LIBRA campaign will achieve higher TBR ($\sim$1) by placing the neutron generator inside the salt volume, achieving higher solid angle coverage.
 
At a volume of \SI{100}{\milli\liter}, the \gls{tbr} values for FLiBe and ClLiF are comparable.
Due to the purely fast neutron flux in the crucible, Li-7 is the dominant tritium breeding isotope, with a tritium yield at least one order of magnitude higher than Li-6.
At these energies Fluorine has a nonnegligible contribution to the \gls{tbr} of all molten salts, as well as beryllium in the case of FLiBe.

In our study, we observed that the measured \gls{tbr} fell significantly short of the predictions made by the OpenMC simulation (see \cref{fig:comparison modelled measured tbr}).
This discrepancy suggests that a substantial portion of the tritium produced during the experiment was not effectively collected.
Since the bubbler collection efficiency was found to be close to \SI{100}{\%}, we hypothesize that the loss of tritium through permeation across the container walls could explain this observed discrepancy.
This phenomenon is further elucidated in the following section, where we discuss the transient tritium release model in detail.

\subsection{Transient tritium release}\label{transient model}

\begin{figure*}[h!]
     \centering
     \begin{subfigure}[b]{0.45\textwidth}
         \centering
         \includegraphics[width=1\linewidth]{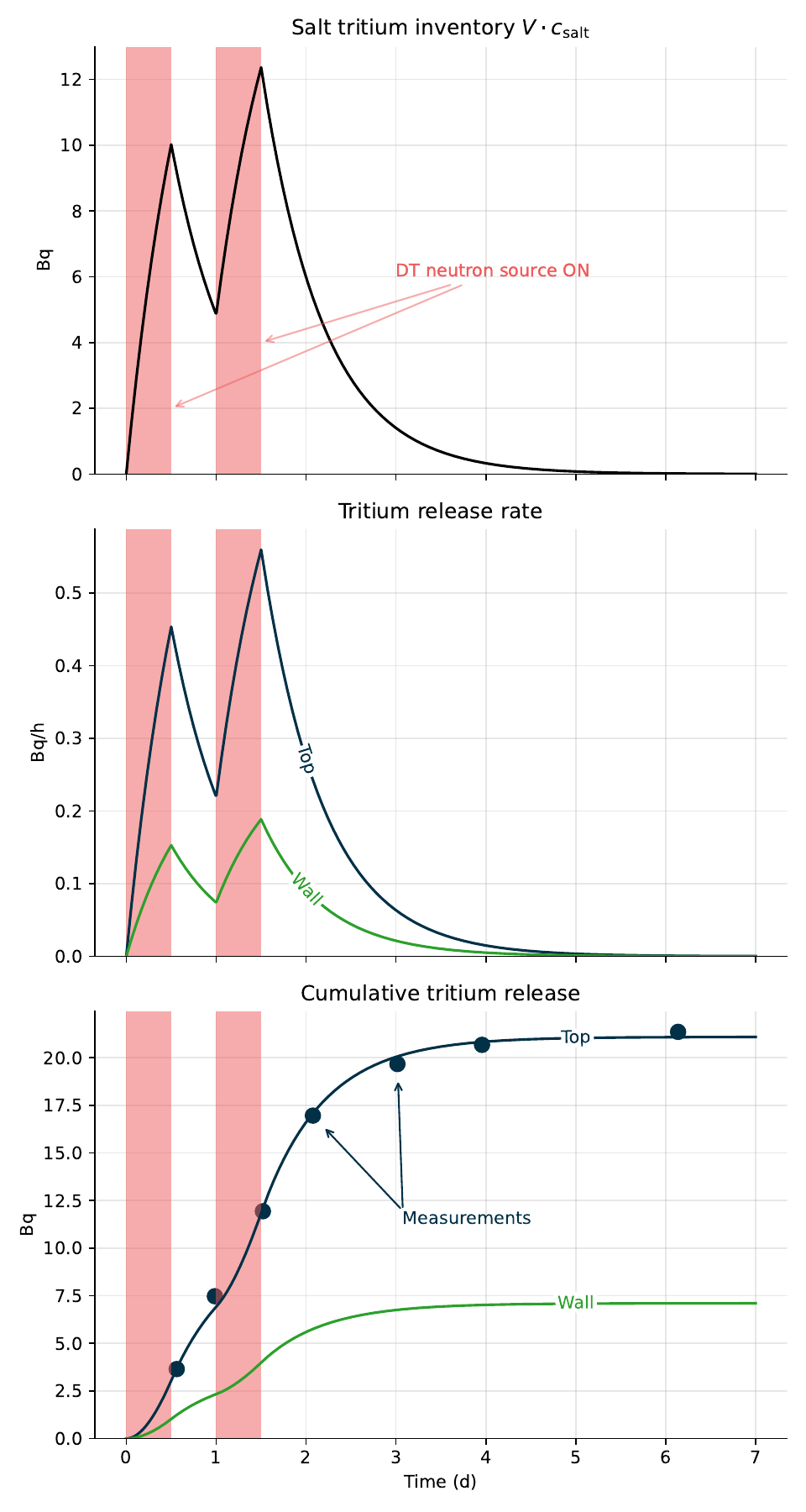}
         \caption{Modeling of the temporal evolution of the tritium salt inventory $V \cdot c_\mathrm{salt}$ (top), tritium release rates $Q_i$ (middle), and cumulative tritium release $\int_0^t Q_i dt $ with comparison to measurements (bottom).}
         \label{fig:model results}
     \end{subfigure}%
     \hfill
     \begin{subfigure}[b]{0.5\textwidth}
        \centering
        \begin{subfigure}[b]{1\textwidth}
            \centering
            \begin{overpic}[width=1\linewidth]{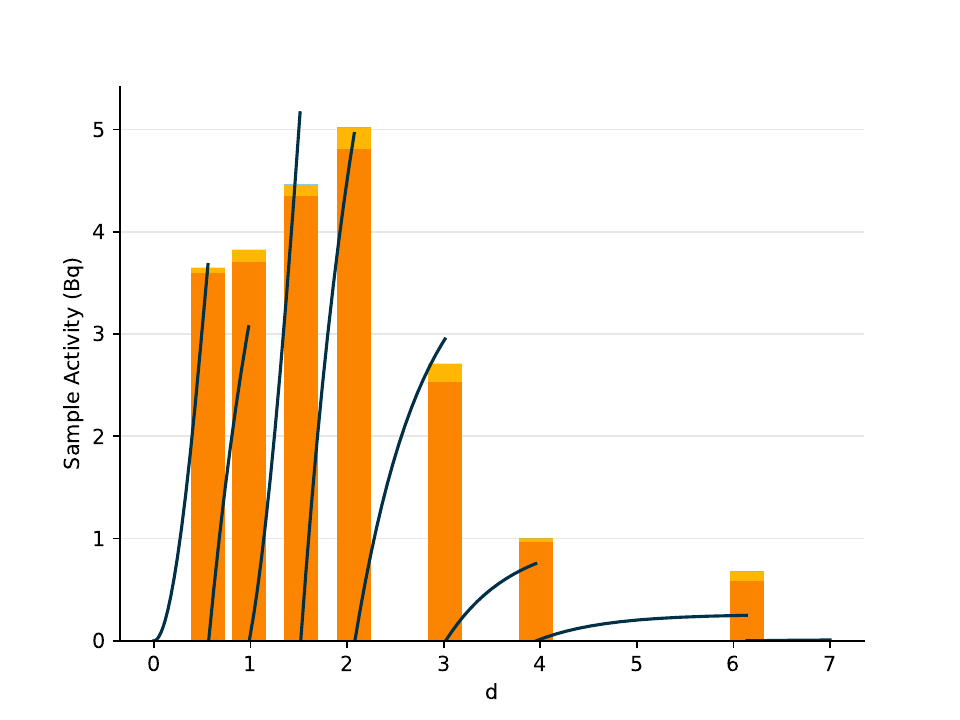}
            \put (50, 50) {\includegraphics[width=0.5\linewidth]{figures/Bubbler_sketch.pdf}}
            \end{overpic}
            \caption{Comparison of the measured (bars) water activity with the modeled (line) sample activity. No measurable amounts of soluble forms of tritium were found in vials \#1 and \#2.}
            \label{fig:sample activity}
        \end{subfigure}
        \begin{subfigure}[b]{1\textwidth}
            \centering
            \includegraphics[width=1\linewidth]{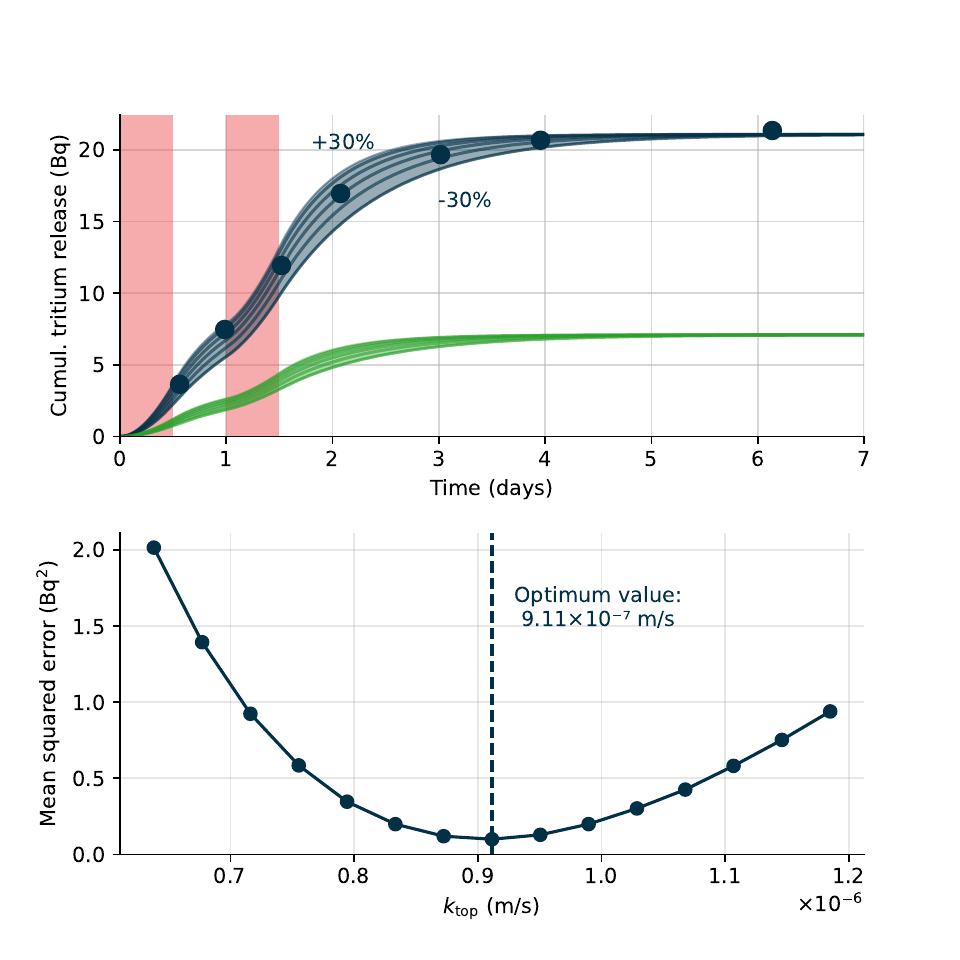}
            \caption{Temporal evolution of the cumulative tritium release (top) and evolution of the mean squared error (bottom) for different values of $k_\mathrm{top}$. $k_\mathrm{wall}$ was varied accordingly.}
            \label{fig:sensitivity study}
        \end{subfigure}
     \end{subfigure}

    \caption{Comparison of the transient tritium release model, which predicts tritium release through the crucible walls and from the top surface of the salt inside the crucible, with the tritium release measurements (here, only tritium release from the top surface of the salt was measured experimentally). The red-shaded regions identify the irradiation periods.}
    \label{fig:tritium results}
\end{figure*}

Following the modeling strategy proposed by Kumagai \textit{et al.} \cite{kumagai_tritium_2018}, the temporal evolution of the salt tritium concentration $c_\mathrm{salt}$ is described by:

\begin{equation}
    V \frac{d c_\mathrm{salt}}{dt} = S - Q_\mathrm{wall} - Q_\mathrm{top} \label{eq:tritium ode}
\end{equation}
where $V$ is the salt volume and $S$ is the source of tritium in \si{T\per\second} expressed by:

\begin{equation}
    S = \mathrm{TBR} \cdot \Gamma_\mathrm{n} 
\end{equation}
where $\Gamma_\mathrm{n} $ is the neutron rate in \si{neutron\per\second}.

The tritium release rates $Q_i$ in \si{T\per\second} are expressed as:
\begin{align}
    Q_i &= A_i \ k_i \ (c_\mathrm{salt} - c_\mathrm{external}) \\
    &\approx A_i \ k_i \ c_\mathrm{salt}
\end{align}
where $A_i$ is the surface area of each release pathway in \si{\metre\squared}, $c_\mathrm{external}$ is the tritium concentration in the gas phase (assumed negligible), and $k_i$ is the mass transport coefficient in \si{\metre\per\second}.

\cref{eq:tritium ode} is solved numerically and the temporal evolution of $c_\mathrm{salt}$ is obtained.
The release rates $Q_i$ can be obtained from $c_\mathrm{salt}$ and the cumulative release can be expressed as $\int_0^t Q_i \ dt$.
This is the quantity measured by the tritium detection system.
The mass transport coefficients $k_i$ are then varied in order to fit the experimental data.

It should be noted that the total production of tritium $\int_0^t S \ dt$ in the salt can be calculated independently of the mass transport coefficients.
Indeed, by integrating \cref{eq:tritium ode}, one can obtain:
\begin{equation}
    V \left[c_\mathrm{salt}(t=t_\mathrm{f}) - c_\mathrm{salt}(t=0) \right] = \int_0^{t_\mathrm{f}} S \ dt - \int_0^{t_\mathrm{f}} \sum_i Q_i \ dt
\end{equation}
where $t_\mathrm{f}$ is the final time.

If $c_\mathrm{salt}(t=t_\mathrm{f}) = c_\mathrm{salt}(t=0) = 0$, which is equivalent to $Q_i(t=t_\mathrm{f}) = Q_i(t=0) = 0$, then:

\begin{equation}
    \int_0^{t_\mathrm{f}} S \ dt = \int_0^{t_\mathrm{f}} \sum_i Q_i \ dt
\end{equation}

Assuming $S$ is constant during irradiation periods:
\begin{align}
    &S \ \Delta t_\mathrm{irradiation} = \int_0^{t_\mathrm{f}} \sum_i Q_i \ dt \\
    &S = \Delta t_\mathrm{irradiation}^{-1} \ \int_0^{t_\mathrm{f}} \sum_i Q_i \ dt
\end{align}
where $\Delta t_\mathrm{irradiation} = \SI{24}{h}$ is the total irradiation time.

In other words, the source term of tritium $S$ can be determined by measuring the cumulative tritium release over the span of the experiment as long as the initial and final concentrations of tritium in the salt  are zero.
Furthermore, the \gls{tbr} can be experimentally measured by a ratio of the total amount of tritium produced to the total amount of neutrons produced.

\begin{table}
    \centering
    \begin{tabular}{ll}
        Parameter & Value \\
        \hline\\
        $V$ & \SI{100}{\centi\metre\cubed} \\
        $A_\mathrm{top}$ & \SI{13.80}{\centi\metre\squared}\\
        $A_\mathrm{wall}$ & \SI{116.18}{\centi\metre\squared}\\
        $k_\mathrm{top}$ & \SI{9.11e-7}{\metre\per\second}\\
        $k_\mathrm{wall}$ & \SI{3.65e-8}{\metre\per\second}\\
        $S$ & \SI{1.83e5}{T\per\second}\\
    \end{tabular}
    \caption{Parameters used in the tritium release model}
    \label{tab:parameters}
\end{table}

The parameters in the tritium release model are presented in \cref{tab:parameters}.
The source $S$ was assumed constant during the irradiation periods (as indicated by the neutron measurements in \cref{tbr_measurement}).
The \gls{tbr} calculated by OpenMC was used as well as the measured neutron rate to determine the tritium source term.
The values of the mass transport coefficient for release of tritium through the top surface of the salt, $k_\mathrm{top}$, and the mass transport coefficient for tritium moving through the crucible walls, $k_\mathrm{wall}$, were optimized to minimize the mean square error with the cumulative release measurements.
The model agrees very well with the experimental data (see \cref{fig:model results} and \cref{fig:sample activity}).
$k_\mathrm{top}$ was determined to be almost twice the value measured by Kumagai \textit{et al} \cite{kumagai_tritium_2018}(\SI{9.11e-7}{\metre\per\s}
 versus \SI{4.9e-7}{\metre\per\s}, respectively).
$k_\mathrm{wall}$ is also almost twice the value measured by Kumagai (\SI{1.9e-8}{\metre\per\s}).
The discrepancies are attributable to the different salt (FLiNaK instead of ClLiF), different geometries, and lower temperature (\SI{500}{\celsius} versus \SI{700}{\celsius} here) used in \cite{kumagai_tritium_2018}.
%The differences can be attributed to the use of different salts at different temperatures (Kumagai \textit{et al} irradiated FLiNaK at \SI{500}{\celsius}) and different geometries.

The Sherwood number can be computed (see \cref{eq:sherwood}) from the measured mass transfer coefficient.
Using the height of the crucible as characteristic length $L$ and the diffusivities of FLiNaK and FLiBe measured in the right temperature range (HTM database v0.14 \cite{delaporte-mathurin_remdelaportemathurinh-transport-materials_2023}), it ranges from 0.8 to 21.6 with an average value of 10.3.
This suggests that convective mass transfer dominates tritium transport in the crucible.
No suitable mass transport correlations were found in the literature for this problem \cite{cussler_diffusion_2009}.
In future work, computational studies will be performed in order to estimate the Sherwood number from a numerical model using the tritium transport code FESTIM \cite{delaporte-mathurin_festim_2024}.

A sensitivity study was performed and the value of $k_\mathrm{top}$ was varied by $\pm \SI{30}{\%}$ from its optimal value ($k_\mathrm{wall}$ was varied by the same factor).
As explained above, the value of the total cumulative release is independent of the mass transport coefficient (see \cref{fig:sensitivity study}).
As the value of the mass transport coefficient increases, the time required to release all tritium from the salt decreases (\textit{i.e.} the characteristic time of the system is shorter).

\subsection{Chemical forms of tritium}\label{speciation}
The entirety of the collected tritium was in non-soluble forms (HT or T$_2$) during the described runs, which used pure He as a sweep gas.
This result differs from previous observations \cite{kumagai_tritium_2018, terai_tritium_2001, nishimura_chemical_2001} in which tritium was found as soluble TF or HTO in FLiNaK and FLiBe when swept with pure He.
Several explanations are possible.
First, we are using a different salt which affects the chemistry of tritium speciation: more oxidizing salts form preferentially produce soluble forms of tritium whereas reducing salts preferentially produce insoluble forms of tritium.
To confirm this, we are planning an experiment in which europium fluoride is added to the salt to make it more oxidizing and expect tritium to be released mainly as TF/TCl.
Another hypothesis was that the potential presence of H$_2$ (residual from the initial conditioning step with \SI{3}{\percent} H$_2$ (see \cref{principle}) led to the formation of HT by isotopic exchange.
We tested this hypothesis by removing this conditioning step, but observed no change in speciation.

\subsection{Reproducibility}

\begin{figure}[h!]
    \centering
    \includegraphics[width=1\linewidth]{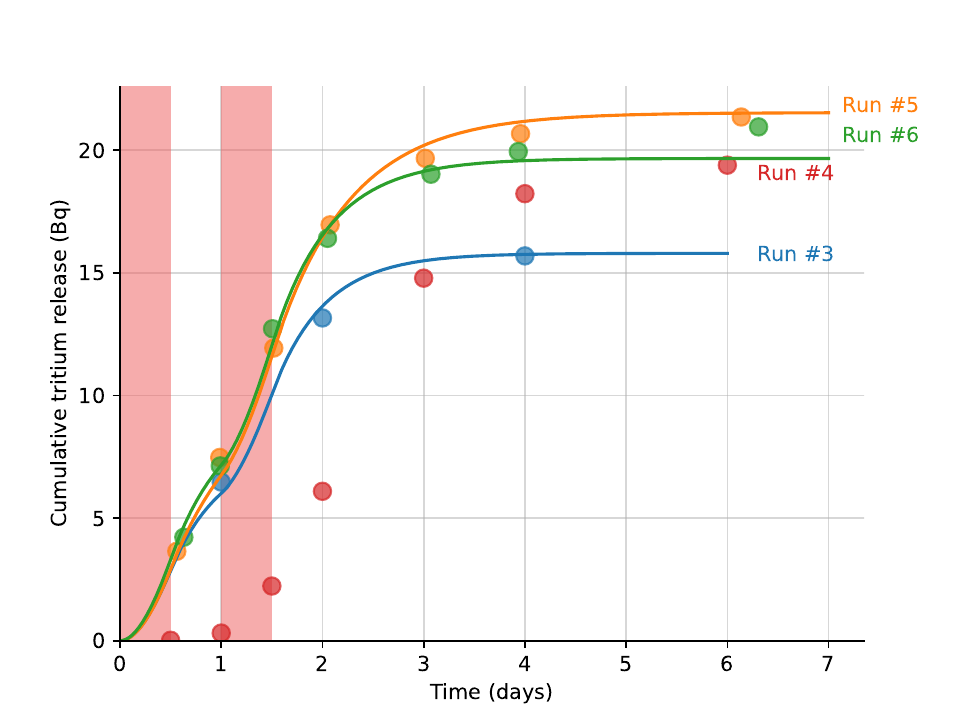}
    \caption{Comparison of the cumulative release measured in runs \#3, \#4, \#5 and \#6. The red-shaded regions identify the irradiation periods. Runs \#1 and \#2 were initial troubleshooting runs during which we did not detect tritium.}
    \label{fig:reproductibility}
\end{figure}

These results have been reproduced in other runs (see \cref{fig:reproductibility}).
For all runs, the plateau value is similar with a relative standard deviation of \SI{12.28}{\%}.
Runs \#1 and \#2 were initial troubleshooting runs during which we did not detect tritium.
Moreover, the neutron rate measurements showed similar results (see \cref{tab:activation_foil_data}).
The dynamics of the release are also very similar with the exception of run \#4 where little tritium was detected before the end of the second irradiation.
While the reasons of this behavior are still unclear, one hypothesis is that tritium was being held up somewhere in the apparatus and got released between day 2 and day 3.
This could explain the large difference between the fourth and fifth data point that is not observable in runs \#3 \#5 and \#6.
The experimental setup was sometimes subject to pipe clogging due to the salt migrating in the tubing and solidifying.
Other reasons could be a change in the convective flow of the salt leading to a reduced mass transport coefficient.

\begin{table}[h]
    \centering
    \begin{tabular}{lll}
       Run  & A-325 & P-383\\
       \hline\\
       1 - 3 & Not measured & Not measured \\
       4 & \SI{1.02e8}{n\per\second}& \SI{2.64e8}{n\per\second}\\
       5 & \SI{1.16e8}{n\per\second}& \SI{2.71e8}{n\per\second}\\
       6 & \SI{1.06e8}{n\per\second}& \SI{2.48e8}{n\per\second}\\
    \end{tabular}
    \caption{Neutron rates measurements from activation foil analysis. A-325 and P-383 refer to the sealed-tube neutron generators described in \cref{neutronsource}.}
    \label{tab:activation_foil_data}
\end{table}
\section{Discussion}

Although these results are extremely encouraging and demonstrate capability in measuring a \gls{tbr} in molten salts, several points will be improved in future experiments.

First, we found that the measured \gls{tbr} was 2.3 times less than the expected \gls{tbr} calculated using OpenMC.
Future iterations of the BABY setup surround the crucible with a second vessel.
Sweep gas through the interstitial space will capture the release of tritium that permeates through the crucible walls.
This will test the hypothesis formulated in \cref{transient model} that there are nonnegligible permeation losses through the crucible walls.
Assuming that the parameterization of the model detailed in this section is correct, the mass transfer coefficient for the wall release is an order of magnitude higher than that calculated by Kumagai \textit{et al.} \cite{kumagai_tritium_2018}.
In one of the experiments, the authors were unable to measure the permeated tritium, as it was below the \gls{lsc} detection limit.
At the high temperatures employed in BABY, the Inconel crucible wall is not expected to act as a tritium permeation barrier.
However, it is possible that the components wrapped around the crucible for heating and insulation reduced the permeability of the system.
The iteration of BABY reported in this work uses alumina as thermal insulation, which is very porous and is not expected to inhibit permeation.
However, future configurations of BABY will attempt to experimentally clarify how much tritium permeates through the walls.

Second, a better characterization of the neutron sources is planned in order to better understand the spatial distribution of the neutron flux and energy spectrum.
This will greatly increase the fidelity of the OpenMC model.

The tritium speciation observed (see \cref{speciation}) does not agree with what was observed in FLiBe or FLiNaK with pure He as a sweep gas \cite{kumagai_tritium_2018, terai_tritium_2001, nishimura_chemical_2001}.
In these experiments, very little to no insoluble tritium (HT, T2) compared to soluble forms of tritium (TF, HTO) was found.
The different redox potentials of the salts are likely to be the reason for this discrepancy and experiments are planned to confirm it.
The presence of H2 impurities could also potentially lead to the formation of HT by isotopic exchange.

\section{Methods}\label{methods}
\subsection{Tritium collection and LSC counting}

Each vial contains \SI{10}{\milli\liter} of deionized water.
When sampling, the tritiated water of each vial is mixed with \SI{10}{\milli\liter} of liquid scintillation cocktail (Ultima Gold\textsuperscript{TM} LLT) and transferred to the \gls{lsc} equipment (model PerkinElmer Tri-Carb 5110 TR) and the total tritium activity in the sample is measured with a counting time of \SI{2}{h}.
The file for the \gls{lsc} method is also available as supplementary data.

\subsection{Activation foil analysis}

Niobium activation foils from Shieldwerx were placed in the same plane as shown in \cref{fig:openmc-model} at the top of each neutron generator for runs 4, 5, and 6.
After each run, the foils were placed between two \qtyproduct[product-units = single]{100 x 100 x 100}{\milli\metre} cubic sodium iodide (NaI) scintillation detectors equipped with photo-multiplier tubes (PMT) and connected to CAEN pulse digitization hardware.  %(make and model numbers)
The pulses from the counting setup were collected using the ADAQAcquisition data acquisition software.
During exposure to the flux of neutrons from the generators, activation occurs through the \mbox{\textsuperscript{93}Nb(n, 2n)\textsuperscript{92m}Nb} reaction.
The de-excitation of \textsuperscript{92m}Nb results in the emission of gamma rays with energy of \SI{934.4}{keV} with a half-life of 10.25 days \cite{jednorog_JETNAA_2017}.
The efficiency of detecting \SI{934.4}{keV} photons was inferred from the measurement of the intrinsic efficiency of the detectors with \SI{661.7}{keV} photons from \textsuperscript{137}Cs decay and \SI{511}{keV} and \SI{1274.5}{keV} photons from \textsuperscript{22}Na decay.

The time history of the activation foil from irradiation through the final count of gammas emitted from \textsuperscript{92m}Nb are used to calculate the average flux of neutrons the foil experienced during the two irradiation periods.
Once the neutron flux in \si{n\per\second\per\centi\meter\squared} is calculated, we assume an isotropic distribution to obtain the neutron rate in \si{n\per\second} based on the position of the foil.

Complete description of the post-process of the activation foil analysis is available as supplementary data.

\subsection{Diamond detector}

The Cividec diamond detector (Cx-L Spectroscopic Amplifier) was placed approximately \SI{35}{\centi\meter} away from the center of the crucible and perpendicular to the target plane of both neutron generators.
Additionally, it was also placed approximately level with the radial center of the A-325 neutron generator.
The detector held a bias voltage of \SI{400}{\volt}, as recommended by the manufacturer.

Detector pulses arising from incident neutrons reacting with the diamond lattice were counted using the CAEN CoMPASS software, thus providing a time-dependent count rate, proportional to the sum of the neutron rates from the two generators.

\subsection{Neutronics OpenMC model}

The (n,Xt) reactions were tallied in the salt region.
The neutron generators were modeled as point sources placed on the tritium-based target position.
The energy and angular distribution were modeled based on a characterization carried out on the A-325 generator, assuming that the P-383 had similar emission characteristics.
The Fusion Evaluated Nuclear Data Library (FENDL-3.1d) library was used for cross sections data.
OpenMC v0.14.0 was used.
The complete materials composition in the model can be found as supplementary data.
Particular attention has been given to the modeling of the neutron sources.
The angular and energy distributions of the neutron source A-325 were previously characterized using a diamond detector.
We model it as a point source with energy and azimuthal angular distributions based on measurements, assuming axial symmetry in the polar direction.
The same distributions are assumed for the P-383 source. 

\subsection{Geometry}

The detailed geometry is available as supplementary data.

\section{Conclusion}

The \gls{baby} experiment represents a significant step forward in the quest for tritium self-sufficiency in fusion power plants.
Despite its preliminary nature, the findings offer valuable insights into the intricate dynamics of tritium breeding within molten salts.
A \gls{tbr} of \num{3.57e-4} was measured: \SI{21}{\becquerel} of tritium produced and a neutron fluence of \SI{3.35e13}{n} over \SI{24}{h} of irradiation.
Future endeavors will be dedicated to reconciling disparities between observed and modeled tritium breeding ratios, refining neutron source characterization, and better understanding of the tritium speciation observations.
Moreover, an effort will be made to develop and validate higher-fidelity tritium transport models.
This project will progress towards larger-scale investigations, such as the planned \SI{1}{\liter} volume experiment with FLiBe salt, and will continue to explore alternative beryllium-free molten salt breeders.

\section*{Acknowledgements}
The authors would like to acknowledge funding from the Advanced Research Projects Agency-Energy (ARPA-E), U.S. Department of Energy.
The views and opinions of authors expressed herein do not necessarily state or reflect those of the United States Government or any agency thereof.

\bibliography{references.bib}

% Generated by IEEEtran.bst, version: 1.14 (2015/08/26)
\begin{thebibliography}{10}
\providecommand{\url}[1]{#1}
\csname url@samestyle\endcsname
\providecommand{\newblock}{\relax}
\providecommand{\bibinfo}[2]{#2}
\providecommand{\BIBentrySTDinterwordspacing}{\spaceskip=0pt\relax}
\providecommand{\BIBentryALTinterwordstretchfactor}{4}
\providecommand{\BIBentryALTinterwordspacing}{\spaceskip=\fontdimen2\font plus
\BIBentryALTinterwordstretchfactor\fontdimen3\font minus \fontdimen4\font\relax}
\providecommand{\BIBforeignlanguage}[2]{{%
\expandafter\ifx\csname l@#1\endcsname\relax
\typeout{** WARNING: IEEEtran.bst: No hyphenation pattern has been}%
\typeout{** loaded for the language `#1'. Using the pattern for}%
\typeout{** the default language instead.}%
\else
\language=\csname l@#1\endcsname
\fi
#2}}
\providecommand{\BIBdecl}{\relax}
\BIBdecl

\bibitem{fia2023}
{Fusion Industry Association}, \emph{The global fusion industry in 2023}, 2023, available at: \url{hhttps://www.fusionindustryassociation.org/wp-content/uploads/2023/07/FIA%E2%80%932023-FINAL.pdf}. Accessed: 2024-21-03.

\bibitem{kovari_tritium_2017}
\BIBentryALTinterwordspacing
M.~Kovari, M.~Coleman, I.~Cristescu, and R.~Smith, ``\BIBforeignlanguage{en}{Tritium resources available for fusion reactors},'' \emph{\BIBforeignlanguage{en}{Nuclear Fusion}}, vol.~58, no.~2, p. 026010, Dec. 2017, publisher: IOP Publishing. [Online]. Available: \url{https://dx.doi.org/10.1088/1741-4326/aa9d25}
\BIBentrySTDinterwordspacing

\bibitem{pearson_tritium_2018}
\BIBentryALTinterwordspacing
R.~J. Pearson, A.~B. Antoniazzi, and W.~J. Nuttall, ``Tritium supply and use: a key issue for the development of nuclear fusion energy,'' \emph{Fusion Engineering and Design}, vol. 136, pp. 1140--1148, Nov. 2018. [Online]. Available: \url{https://www.sciencedirect.com/science/article/pii/S092037961830379X}
\BIBentrySTDinterwordspacing

\bibitem{coleman_demo_2019}
\BIBentryALTinterwordspacing
M.~Coleman, Y.~Hörstensmeyer, and F.~Cismondi, ``{DEMO} tritium fuel cycle: performance, parameter explorations, and design space constraints,'' \emph{Fusion Engineering and Design}, vol. 141, pp. 79--90, Apr. 2019. [Online]. Available: \url{https://www.sciencedirect.com/science/article/pii/S092037961930167X}
\BIBentrySTDinterwordspacing

\bibitem{chen_tritium_2016}
\BIBentryALTinterwordspacing
H.~Chen, L.~Pan, Z.~Lv, W.~Li, and Q.~Zeng, ``Tritium fuel cycle modeling and tritium breeding analysis for {CFETR},'' \emph{Fusion Engineering and Design}, vol. 106, pp. 17--20, May 2016. [Online]. Available: \url{https://www.sciencedirect.com/science/article/pii/S0920379616301922}
\BIBentrySTDinterwordspacing

\bibitem{meschini_modeling_2023}
\BIBentryALTinterwordspacing
S.~Meschini, S.~E. Ferry, R.~Delaporte-Mathurin, and D.~G. Whyte, ``\BIBforeignlanguage{en}{Modeling and analysis of the tritium fuel cycle for {ARC}- and {STEP}-class {D}-{T} fusion power plants},'' \emph{\BIBforeignlanguage{en}{Nuclear Fusion}}, vol.~63, no.~12, p. 126005, Sep. 2023, publisher: IOP Publishing. [Online]. Available: \url{https://dx.doi.org/10.1088/1741-4326/acf3fc}
\BIBentrySTDinterwordspacing

\bibitem{abdou_physics_2021}
\BIBentryALTinterwordspacing
M.~Abdou, M.~Riva, A.~Ying, C.~Day, A.~Loarte, L.~Baylor, P.~Humrickhouse, T.~F. Fuerst, and S.~Cho, ``\BIBforeignlanguage{en}{Physics and technology considerations for the deuterium–tritium fuel cycle and conditions for tritium fuel self sufficiency},'' \emph{\BIBforeignlanguage{en}{Nuclear Fusion}}, vol.~61, no.~1, p. 013001, Jan. 2021. [Online]. Available: \url{https://iopscience.iop.org/article/10.1088/1741-4326/abbf35}
\BIBentrySTDinterwordspacing

\bibitem{ferry_libra_2022}
\BIBentryALTinterwordspacing
S.~E. Ferry, K.~B. Woller, E.~E. Peterson, C.~Sorensen, and D.~G. Whyte, ``The {LIBRA} {Experiment}: {Investigating} {Robust} {Tritium} {Accountancy} in {Molten} {FLiBe} {Exposed} to a {D}-{T} {Fusion} {Neutron} {Spectrum},'' \emph{Fusion Science and Technology}, vol.~0, no.~0, pp. 1--23, Jun. 2022, publisher: Taylor \& Francis \_eprint: https://doi.org/10.1080/15361055.2022.2078136. [Online]. Available: \url{https://doi.org/10.1080/15361055.2022.2078136}
\BIBentrySTDinterwordspacing

\bibitem{sorbom_arc_2015}
\BIBentryALTinterwordspacing
B.~N. Sorbom, J.~Ball, T.~R. Palmer, F.~J. Mangiarotti, J.~M. Sierchio, P.~Bonoli, C.~Kasten, D.~A. Sutherland, H.~S. Barnard, C.~B. Haakonsen, J.~Goh, C.~Sung, and D.~G. Whyte, ``\BIBforeignlanguage{en}{{ARC}: {A} compact, high-field, fusion nuclear science facility and demonstration power plant with demountable magnets},'' \emph{\BIBforeignlanguage{en}{Fusion Engineering and Design}}, vol. 100, pp. 378--405, Nov. 2015. [Online]. Available: \url{http://www.sciencedirect.com/science/article/pii/S0920379615302337}
\BIBentrySTDinterwordspacing

\bibitem{kondo_606_2020}
\BIBentryALTinterwordspacing
M.~Kondo, T.~Tanaka, S.~Fukada, and T.~Valentyn, ``\BIBforeignlanguage{en}{6.06 - {Liquid} {Breeder} {Materials}},'' in \emph{\BIBforeignlanguage{en}{Comprehensive {Nuclear} {Materials} ({Second} {Edition})}}, R.~J.~M. Konings and R.~E. Stoller, Eds.\hskip 1em plus 0.5em minus 0.4em\relax Oxford: Elsevier, Jan. 2020, pp. 176--202. [Online]. Available: \url{https://www.sciencedirect.com/science/article/pii/B9780128035818116194}
\BIBentrySTDinterwordspacing

\bibitem{kuang_conceptual_2018}
\BIBentryALTinterwordspacing
A.~Kuang, N.~Cao, A.~Creely, C.~Dennett, J.~Hecla, B.~LaBombard, R.~Tinguely, E.~Tolman, H.~Hoffman, M.~Major, J.~Ruiz~Ruiz, D.~Brunner, P.~Grover, C.~Laughman, B.~Sorbom, and D.~Whyte, ``\BIBforeignlanguage{en}{Conceptual design study for heat exhaust management in the {ARC} fusion pilot plant},'' \emph{\BIBforeignlanguage{en}{Fusion Engineering and Design}}, vol. 137, pp. 221--242, Dec. 2018. [Online]. Available: \url{https://linkinghub.elsevier.com/retrieve/pii/S0920379618306185}
\BIBentrySTDinterwordspacing

\bibitem{ferrero2023exploration}
G.~Ferrero, S.~Meschini, R.~Testoni, and M.~Zucchetti, ``Exploration of arc-class reactor vessel and divertor cooling system,'' \emph{Fusion Engineering and Design}, vol. 192, p. 113818, 2023.

\bibitem{segantin_optimization_2020}
\BIBentryALTinterwordspacing
S.~Segantin, R.~Testoni, Z.~Hartwig, D.~Whyte, and M.~Zucchetti, ``Optimization of tritium breeding ratio in {ARC} reactor,'' \emph{Fusion Engineering and Design}, vol. 154, p. 111531, May 2020. [Online]. Available: \url{https://www.sciencedirect.com/science/article/pii/S092037962030079X}
\BIBentrySTDinterwordspacing

\bibitem{giancarli2006breeding}
L.~Giancarli, V.~Chuyanov, M.~Abdou, M.~Akiba, B.~Hong, R.~L{\"a}sser, C.~Pan, Y.~Strebkov \emph{et~al.}, ``Breeding blanket modules testing in iter: an international program on the way to demo,'' \emph{Fusion Engineering and Design}, vol.~81, no. 1-7, pp. 393--405, 2006.

\bibitem{forest2020test}
L.~Forest, L.~V. Boccaccini, L.~Cogneau, A.~L. Puma, H.~Neuberger, S.~Pascal, J.~Rey, N.~Thomas, J.~Tosi, and M.~Zmitko, ``Test blanket modules (iter) and breeding blanket (demo): History of major fabrication technologies development of hcll and hcpb and status,'' \emph{Fusion Engineering and Design}, vol. 154, p. 111493, 2020.

\bibitem{wang2019current}
X.~Wang, K.~Feng, Y.~Chen, L.~Zhang, Y.~Feng, X.~Wu, H.~Liao, X.~Ye, F.~Zhao, Q.~Cao \emph{et~al.}, ``Current design and r\&d progress of the chinese helium cooled ceramic breeder test blanket system,'' \emph{Nuclear Fusion}, vol.~59, no.~7, p. 076019, 2019.

\bibitem{hirose2024functional}
T.~Hirose, W.~Guan, T.~Katagiri, A.~Wakasa, Y.~Someya, M.~Nakajima, Y.~Koga, Y.~Miyoshi, T.~Nozawa, Y.~Kawamura \emph{et~al.}, ``Functional tests for water cooled ceramic breeder blanket system using full-scale mockups,'' \emph{Fusion Engineering and Design}, vol. 200, p. 114227, 2024.

\bibitem{batistoni2012neutronics}
P.~Batistoni, M.~Angelone, U.~Fischer, A.~Klix, I.~Kodeli, D.~Leichtle, M.~Pillon, W.~Pohorecki, and R.~Villari, ``Neutronics experiments for uncertainty assessment of tritium breeding in hcpb and hcll blanket mock-ups irradiated with 14 mev neutrons,'' \emph{Nuclear Fusion}, vol.~52, no.~8, p. 083014, 2012.

\bibitem{ochiai2014dt}
K.~Ochiai, Y.~Kawamura, T.~Hoshino, Y.~Edao, K.~Takakura, M.~Ohta, S.~Sato, and C.~Konno, ``Dt neutron irradiation experiment for evaluation of tritium recovery from wccb blanket,'' \emph{Fusion Engineering and Design}, vol.~89, no. 7-8, pp. 1464--1468, 2014.

\bibitem{zhu2021experimental}
Q.~Zhu, W.~Chen, J.~Bao, H.~Du, S.~Liu, and K.~Huang, ``Experimental study on tritium breeding in water-cooled ceramic breeder blanket mock-up under d--t neutron irradiation conditions,'' \emph{Nuclear Fusion}, vol.~61, no.~6, p. 066018, 2021.

\bibitem{humrickhouse_tritium_2020}
\BIBentryALTinterwordspacing
P.~W. Humrickhouse and T.~F. Fuerst, ``\BIBforeignlanguage{English}{Tritium {Transport} {Phenomena} in {Molten}-{Salt} {Reactors}},'' Idaho National Lab. (INL), Idaho Falls, ID (United States), Tech. Rep. INL/EXT-20-59927-Rev000, Sep. 2020. [Online]. Available: \url{https://www.osti.gov/biblio/1777267-tritium-transport-phenomena-molten-salt-reactors}
\BIBentrySTDinterwordspacing

\bibitem{greenberg_neutron_2011}
\BIBentryALTinterwordspacing
R.~R. Greenberg, P.~Bode, and E.~A. De~Nadai~Fernandes, ``Neutron activation analysis: {A} primary method of measurement,'' \emph{Spectrochimica Acta Part B: Atomic Spectroscopy}, vol.~66, no.~3, pp. 193--241, Mar. 2011. [Online]. Available: \url{https://www.sciencedirect.com/science/article/pii/S0584854711000024}
\BIBentrySTDinterwordspacing

\bibitem{angelone_properties_2021}
\BIBentryALTinterwordspacing
M.~Angelone and C.~Verona, ``\BIBforeignlanguage{en}{Properties of {Diamond}-{Based} {Neutron} {Detectors} {Operated} in {Harsh} {Environments}},'' \emph{\BIBforeignlanguage{en}{Journal of Nuclear Engineering}}, vol.~2, no.~4, pp. 422--470, Dec. 2021, number: 4 Publisher: Multidisciplinary Digital Publishing Institute. [Online]. Available: \url{https://www.mdpi.com/2673-4362/2/4/32}
\BIBentrySTDinterwordspacing

\bibitem{parker_radiometric_2023}
\BIBentryALTinterwordspacing
A.~J. Parker, M.~D. Aspinall, C.~Boxall, F.~D. Cave, and M.~J. Joyce, ``Radiometric techniques for the detection and assessment of tritium in aqueous media - a review,'' \emph{Progress in Nuclear Energy}, vol. 162, p. 104733, Aug. 2023. [Online]. Available: \url{https://www.sciencedirect.com/science/article/pii/S0149197023001683}
\BIBentrySTDinterwordspacing

\bibitem{delaporte-mathurin_libra-projectinsights--baby-experiment-paper_2024}
\BIBentryALTinterwordspacing
R.~Delaporte-Mathurin and S.~Segantin, ``{LIBRA}-project/insights-from-{BABY}-experiment-paper: {Initial} release for submission,'' Apr. 2024. [Online]. Available: \url{https://doi.org/10.5281/zenodo.10904452}
\BIBentrySTDinterwordspacing

\bibitem{romano_openmc_2015}
\BIBentryALTinterwordspacing
P.~K. Romano, N.~E. Horelik, B.~R. Herman, A.~G. Nelson, B.~Forget, and K.~Smith, ``\BIBforeignlanguage{en}{{OpenMC}: {A} state-of-the-art {Monte} {Carlo} code for research and development},'' \emph{\BIBforeignlanguage{en}{Annals of Nuclear Energy}}, vol.~82, pp. 90--97, Aug. 2015. [Online]. Available: \url{https://www.sciencedirect.com/science/article/pii/S030645491400379X}
\BIBentrySTDinterwordspacing

\bibitem{kumagai_tritium_2018}
\BIBentryALTinterwordspacing
K.~Kumagai, T.~Tanaka, J.~Yagi, T.~Watanabe, F.~Sato, S.~Tamaki, I.~Murata, and A.~Sagara, ``Tritium release from molten salt {FLiNaK} under low flux neutron irradiation with an {AmBe} neutron source,'' \emph{Fusion Engineering and Design}, vol. 136, pp. 1269--1272, Nov. 2018. [Online]. Available: \url{https://www.sciencedirect.com/science/article/pii/S0920379618304095}
\BIBentrySTDinterwordspacing

\bibitem{delaporte-mathurin_remdelaportemathurinh-transport-materials_2023}
\BIBentryALTinterwordspacing
R.~Delaporte-Mathurin, J.~Dark, natethegreatINL, and T.~Fuerst, ``{RemDelaporteMathurin}/h-transport-materials: {Release} 0.14,'' Aug. 2023. [Online]. Available: \url{https://zenodo.org/records/8256349}
\BIBentrySTDinterwordspacing

\bibitem{cussler_diffusion_2009}
E.~L. Cussler, \emph{\BIBforeignlanguage{en}{Diffusion: {Mass} {Transfer} in {Fluid} {Systems}}}.\hskip 1em plus 0.5em minus 0.4em\relax Cambridge University Press, Jan. 2009, google-Books-ID: dq6LdJyN8ScC.

\bibitem{delaporte-mathurin_festim_2024}
\BIBentryALTinterwordspacing
R.~Delaporte-Mathurin, J.~Dark, G.~Ferrero, E.~A. Hodille, V.~Kulagin, and S.~Meschini, ``{FESTIM}: {An} open-source code for hydrogen transport simulations,'' \emph{International Journal of Hydrogen Energy}, vol.~63, pp. 786--802, Apr. 2024. [Online]. Available: \url{https://www.sciencedirect.com/science/article/pii/S0360319924010218}
\BIBentrySTDinterwordspacing

\bibitem{terai_tritium_2001}
T.~Terai, A.~Suzuki, and S.~Tanaka, ``Tritium release from {Li2BeF4} molten salt breeder under neutron irradiation at elevated temperature,'' \emph{Fusion technology}, vol.~39, no. 2P2, pp. 768--772, 2001, publisher: Taylor \& Francis.

\bibitem{nishimura_chemical_2001}
\BIBentryALTinterwordspacing
H.~Nishimura, A.~Suzuki, T.~Terai, M.~Yamawaki, S.~Tanaka, A.~Sagara, and O.~Motojima, ``\BIBforeignlanguage{en}{Chemical behavior of {Li2BeF4} molten salt as a liquid tritium breeder},'' \emph{\BIBforeignlanguage{en}{Fusion Engineering and Design}}, vol. 58-59, pp. 667--672, Nov. 2001. [Online]. Available: \url{http://www.sciencedirect.com/science/article/pii/S0920379601005816}
\BIBentrySTDinterwordspacing

\bibitem{jednorog_JETNAA_2017}
\BIBentryALTinterwordspacing
J.~S, L.~E, B.~P, B.~B, C.~A, G.~Z, G.~L, K.~A, L.~S, M.~K, P.~L, P.~A, P.~M, P.~S, R.~M, R.~D, R.~N, T.~M, T.~D, and J.~Contributors, ``\BIBforeignlanguage{en}{Activation measurements in support of the 14 mev neutron calibration of jet neutron monitors},'' \emph{\BIBforeignlanguage{en}{Fusion Engineering and Design}}, vol. 125, pp. 50--56, 2017. [Online]. Available: \url{https://www.sciencedirect.com/science/article/pii/S0920379617308918}
\BIBentrySTDinterwordspacing

\end{thebibliography}
\end{document}